\documentclass[reprint,nofootinbib,amsmath,amssymb,fontenc aps]{revtex4-1}
\usepackage{graphicx,dcolumn,bm,hyperref,float}
\usepackage{anyfontsize}
\usepackage{color}
\usepackage{xcolor}

\newcommand{\beq}{\begin{equation}}
\newcommand{\eeq}{\end{equation}}
\newcommand{\bea}{\begin{eqnarray}}
\newcommand{\eea}{\end{eqnarray}}
\newcommand{\bef}{\begin{figure}}
\newcommand{\eef}{\end{figure}}

\newcommand{\Phimpp}{\Phi_{\mbox{\tiny{MPP}}}}
\newcommand{\pt}{\phi}
\newcommand{\ptt}{\tilde{\phi}}
\newcommand{\CM}{\mbox{\tiny{CM}}}

\begin{document}

\title{Analysis of anomalous contribution to galactic rotation curves due to stochastic spacetime}
\title{Critical Analysis of Replacing Dark Matter and Dark Energy with a Model of Stochastic Spacetime}

\author{Mark P.~Hertzberg$^{1,2,3}$}
\email{mark.hertzberg@tufts.edu}
\author{Abraham Loeb$^2$}
\email{aloeb@cfa.harvard.edu}
\affiliation{$^1$Institute of Cosmology, Department of Physics and Astronomy, Tufts University, Medford, MA 02155, USA
\looseness=-1}
\affiliation{$^2$Institute of Theory and Computation, Center for Astrophysics, Harvard University, 60 Garden Street, Cambridge, MA 02138, USA
\looseness=-1}
\affiliation{$^3$Center for Theoretical Physics, Department of Physics, 
Massachusetts Institute of Technology, Cambridge, MA 02139, USA}

\begin{abstract}
We analyze consequences of trying to replace dark matter and dark energy with models of stochastic spacetime. In particular, 
we analyze the model put forth by Ref.~\cite{Oppenheim:2024rcp}, in which it is claimed that ``post-quantum classical gravity" (PQCG), a stochastic theory of gravity, leads to modified Newtonian dynamics (MOND) behavior on galactic scales that reproduces galactic rotation curves, and leads to dark energy. We show that this analysis has four basic problems: (i) the equations of PQCG do not lead to a new large scale force of the form claimed in the paper, (ii) the form claimed is not of the MONDian form anyhow and so does not correspond to observed galactic dynamics, (iii) the spectrum of fluctuations is very different from observations, and (iv) we also identify some theoretical problems in these models.
\end{abstract}

\maketitle

%\tableofcontents

\section{Introduction}
%{\em Introduction}.---

An outstanding problem in modern physics is the successful unification of quantum mechanics and gravity. An interesting approach to this problem has been put forward in Refs.~\cite{Oppenheim:2018igd,Oppenheim:2023izn,Layton:2023oud} in which matter is treated quantum-mechanically, while gravity is treated classically; this is dubbed ``post-quantum classical gravity" (PQCG). The coupling is such that gravity becomes effectively stochastic. Whether this framework truly leads to an internally consistent theory is not the primary focus of this paper, though we do identify some theoretical problems in this work also.

What is relevant to this paper is the possibility of such theories leading to large scale testable predictions. Very interestingly, in Ref.~\cite{Oppenheim:2024rcp} it was claimed that PQCG indeed does so, namely that it leads to a new long range force between matter of the modified Newtonian dynamics (MOND) form; the form that can reproduce galactic rotation curves \cite{Milgrom:1983ca,Milgrom:1983pn,Milgrom:1983zz,Bekenstein:1984tv}. While it would be very interesting if the unification of gravity and quantum mechanics leads to such large scale effects. Here we point out that the PQCG theory of the form presented in \cite{Oppenheim:2024rcp} does {\em not} in fact lead to anything like MONDian dynamics. Furthermore, we show that the theory has a very different spectrum of fluctuations than that observed. Whether some other variation of this framework improves upon this is beyond the scope of this paper.

\section{The Newtonian Limit of PQCG  }\label{NewtonianLimit}
%{\em Anisotropic Universe}.---

The full PQCG theoretical framework is an interesting theory in which quantum dynamics of matter and classical dynamics of gravity are coupled together in a novel way. Nevertheless, the dynamics can be encoded in an action \cite{Oppenheim:2023izn}.
The full relativistic theory is somewhat complicated. However, for the purpose of studying galactic dynamics, we only need to pay attention to the low velocity limit of the theory as the characteristic speeds of gas, stars, and satellites in a galaxy are orders of magnitude slower than the speed of light (we set $c=1$). In Ref.~\cite{Layton:2023oud}, the low velocity limit of the theory is explained to be given by the following effective action (see Section \ref{Temporal} for some discussion of relativistic corrections)
\beq
\mathcal{I}= -\alpha\! \int \!dt\, d^3x\left(\nabla^2\Phi-4\pi G_N\rho(x)\right)^2
\label{Action}\eeq
(with additional contributions for the matter degrees of freedom), 
where $\Phi$ is the gravitational potential, $\rho$ is the matter mass density, and $\alpha>0$ is a constant pre-factor. This action determines the evolution of a probability distribution $\varrho$ for the the gravitational field $\Phi$ through a path integral whose integrand is weighted by a factor $\propto \exp(\mathcal{I})$. So configurations that {\em maximize} this action $\mathcal{I}$ can dominate the space of paths; we shall refer to these as the ``most probable paths" (MPPs) (in the literature, it is sometimes just a ``typical" path that is called a MPP accounting for fluctuations around the mean; we return to this later). The corresponding gravitational potential is denoted $\Phimpp$. Stochastic fluctuations around this are exponentially suppressed, depending on the magnitude of the dimensionless constant $\alpha$ (see ahead to Section \ref{FullDistn} for the issue of absorbing a temporal factor into $\alpha$ to make this more precise).
If $\alpha$ is sufficiently large, then we can ignore such fluctuations. However, if $\alpha$ is sufficiently small, we cannot; this latter case shall be analyzed in Section \ref{Fluctuations}.

If no  boundary conditions are specified, then the most probable path simply minimizes the factor in brackets in Eq.~(\ref{Action}). This gives the standard Newtonian potential $\Phimpp=\Phi_N$ obeying the Poisson equation
\beq
\nabla^2\Phi_N=4\pi G_N\rho(x)
\eeq
However, if one specifies boundary conditions on $\Phi$ and $\nabla\Phi$ (both because the action is 4th order in derivatives), then one is not guaranteed to be able to satisfy the Poisson equation.
In this case, the extremal path arises from extremizing Eq.~(\ref{Action}). The corresponding Euler-Lagrange variation readily leads to the equation for $\Phimpp$
\beq
\nabla^4\Phimpp=4\pi G_N\nabla^2\rho
\label{MPoisson}\eeq
This is evidently just the standard Poisson equation for Newtonian gravity; however, it has an additional Laplacian operator on both sides of the equation; we will refer to it as the modified Newton equation (MNE).

\section{Solution of Modified Poisson Equation with Spherical Symmetry}

For this section and the next, we work under the assumption that the fluctuations are small, leaving us with the MNE as the relevant equation for $\Phi$ assuming some non-trivial boundary conditions are imposed (then we include fluctuations in Section \ref{Fluctuations}). 
The MNE is so similar to the standard Poisson equation for Newtonian gravity that we should expect very similar behavior with only subtle differences. However, Ref.~\cite{Oppenheim:2024rcp} claimed that there are dramatic differences. To unpack this, let us consider a localized source and consider the region outside of the source where $\rho=0$. In this vacuum region of space, the equation reduces to 
\beq
\nabla^4\Phimpp=0\quad\quad(\mbox{in vacuum})
\eeq

This equation has infinitely many solutions. However, as a first step, let us consider {\em spherically symmetric} boundary conditions. There is no obvious reason for this assumption (see Section \ref{Fluctuations} for a more general analysis) but it will be useful to identify some key features.

With the assumption of spherical symmetry, the general solution of this equation away from $r=0$ is
\beq
\Phimpp={\kappa_{-1}\over r}+\kappa_0+\kappa_1\,r+\kappa_2\,r^2
\eeq
where $\kappa_{-1,0,1,2}$ are constants in space, although it is not obvious they should be static in time unless static boundary conditions are imposed (see below for more discussion of time dependence). By comparing to the Newtonian theory, we can easily identify
\beq
\kappa_{-1}=-GM
\eeq
where $M$ is the mass of the source, and so the usual Newtonian solution is readily recovered for small $r$. The $\kappa_0$ term is a constant and has no direct consequences. Let us now turn to the next pair of contributions; neither of these solve the usual Poisson equation in vacuum and hence are of high interest.

\subsection{Quadratic Term}

Let us start by examining the $\kappa_2\,r^2$ term. As Ref.~\cite{Oppenheim:2024rcp} notes, if we write
\beq
\kappa_2=-{\Lambda\over 6}
\eeq
then it can play a similar role to a cosmological constant $\Lambda$. However, this relies upon the important {\em assumption} that is is static in time. Ref.~\cite{Oppenheim:2024rcp} claims that indeed it should be static as this derives from a relativistic theory. This argument is unsatisfying as it in fact depends on the choice of boundary conditions and it is not clear why one would impose such static boundary conditions in an expanding universe; naively it could change on the order the Hubble time or other dynamical timescales in the problem. But we shall not develop this point further here.

In any case, {\em if} $\kappa_2=-\Lambda/6$ is static, its consequences can be understood as follows: For very large $r$  the acceleration is $-\nabla\Phimpp\approx\Lambda\, r\,\hat{r}/3$. By equating this to $\ddot{\bf x}=\ddot{r}\,\hat{r}$, we have the differential equation $\ddot{r}=\Lambda\,r/3$. This simple differential equation has the exponential solution $r\propto\exp(\sqrt{\Lambda/3}\,t)$ as is appropriate for a cosmological constant. While this is amusing, we note that including a cosmological constant within classical general relativity is completely standard. So there is nothing obviously new here. In fact the situation is worse here, as one needs to impose that $\kappa_2$ is static and to impose spherically symmetry boundary conditions. These assumptions are not needed in general relativity, as the space-time invariance of $\Lambda$ is locked in by internal consistency of the 2 degrees of freedom of the graviton and local Poincare symmetry.

\subsection{Linear Term}

Now let us turn to the term of most interest; the $\kappa_1\,r$ term. This is the term that is claimed to be responsible for MONDian dynamics in Ref.~\cite{Oppenheim:2024rcp}, as we discuss in the next section. The presence of a linear term is in fact the first problem in this analysis, as we discuss now. While it is true that $\nabla^4r=0$ away from $r=0$, it is not true at $r=0$ (as already noted in Ref.~\cite{Oppenheim:2024rcp}). Recall that the Laplacian in spherical coordinates is
\beq
\nabla^2=\partial_r^2+{2\over r}\partial_r
\eeq
Let us denote $\Phi_1\equiv \kappa_1\,r$, we then have
\beq
\nabla^2 \Phi_1={2\kappa_1\over r}
\eeq
and in turn we have
\beq
\nabla^4\Phi_1=-8\pi\,\kappa_1\,\delta^3({\bf x})
\eeq
Now, while this reproduces the desired $\nabla^4\Phi=0$ for non-zero $r$, it cannot match onto a localized source. In order to obey the MNE in all of space, one would need a mass density $\rho_1(x)$ that obeys
\beq
G_N\nabla^2\rho_1=-2\kappa_1\,\delta^3({\bf x})
\eeq
This would require a mass density that is itself de-localized as it would need to obey
\beq
\rho_1\propto {1\over r}
\eeq
and hence one would never be in the actual vacuum in the first place. So, in fact, the new linear term $\kappa_1\,r$ is {\em forbidden} when one considers the full solution. Stated differently, the only situation in which there is a linear term for $\Phimpp$ is when there is a $1/r$ mass density profile; but this is already a property of Newtonian gravity anyhow, and so this is not new after all.

Instead, the most general solution of the MNE equation can be written as
\beq
\Phimpp=\Phi_N+\Phi_h
\eeq
where these contributions obey
\beq
\nabla^2\Phi_N=4\pi G_N\rho,\quad\quad \nabla^4\Phi_h=0
\eeq
where $\Phi_N$ is the standard  potential of Newtonian gravity, and $\Phi_h$ obeys the {\em homogeneous} form of the MNE throughout {\em all} space, not just in vacuum. For spherically symmetric configurations, the {\em only} solution for $\Phi_h$ (apart from a constant) is just the quadratic term
\beq
\Phi_h=\kappa_2\,r^2
\eeq
as already discussed above. 
So in this theory, the only new contribution to Newtonian gravity is a cosmological constant term, but the claimed new linear term is $\kappa_1\,r$ is in fact forbidden.

\subsection{Another Derivation}

To be extra careful, let us derive the absence of the linear term from another point of view.
Suppose we take the MNE (\ref{MPoisson}) and integrate it over a ball of radius $R$
\beq
\int_{ball} d^3 x\nabla^4\Phimpp=4\pi G_N\int_{ball} d^3x\nabla^2\rho
\eeq
By the divergence theorem we can write both sides as a boundary integral over the sphere of radius $R$
\beq
\int_{sphere} d^2S\, \partial_r(\nabla^2\Phimpp)=4\pi G_N\int_{sphere} d^2S \,\partial_r\rho
\label{sphere}\eeq
Now if we are in vacuum $\rho=0$ at some finite radius $R$, then the right hand side vanishes. Furthermore if we have spherical symmetry, then the angular integral on the left hand side is just a factor $\int_{sphere} d^2S=4\pi R^2$. This leaves us with the requirement
\beq
\partial_r(\nabla^2\Phimpp)\Big{|}_{r=R}=0\,\,\,(\mbox{in vacuum}, R>0)
\eeq
Again using spherical symmetry, we can re-write this as
\beq
\partial_r(\partial_r^2\Phimpp+{2\over r}\partial_r\Phimpp)\Big{|}_{r=R}=0\,\,\,(\mbox{in vacuum}, R>0)
\label{req}\eeq
The general solution of this 3rd order differential equation is
\beq
\Phimpp(r) = {\kappa_{-1}\over r}+\kappa_0+\kappa_2r^2
\label{gen}\eeq
(replacing $R\to r$ for ease of notation), 
where $\kappa_{-1},\kappa_0,\kappa_2$ are constants. As mentioned above, $\kappa_{-1}=-GM$, $\kappa_0$ is an irrelevant constant in the Newtonian limit, and $\kappa_2$ plays a role akin to the cosmological constant under the assumption that it is static. 

However $\kappa_1r$ does {\em not} solve Eq.~(\ref{req}) in any sense. If we try $\Phi_1=\kappa_1 r$ we obtain
\beq
\partial_r(\partial_r^2\Phi_1+{2\over r}\partial_r\Phi_1)=-{2\kappa_1\over r^2}
\eeq
which in no sense vanishes. Instead we see that this requires $\rho\neq 0$ and so we are not in the vacuum. By returning to Eq.~(\ref{sphere}), we see that this requires
\beq
-{2\kappa_1\over r^2}=4\pi G_N\partial_r\rho
\eeq
Hence we would require
\beq
\rho={\kappa_1\over 2\pi G_N\,r}
\eeq
(up to a constant). So this requires $\rho\propto 1/r$ and so we would definitively {\em not be in the vacuum}; so this is not a black hole solution at all. This confirms the points already made above.

In fact more generally, if we assume a power law $\Phi_p=\kappa_p r^p$, we have the requirement to actually be in the vacuum (away from $r=0$) of
\bea
&&\partial_r(\partial_r^2\Phi_p+{2\over r}\partial_r\Phi_p)=\nonumber\\
&&\kappa_p(p-2)p(p+1)r^{-3+p} = 0 \,\,\,(\mbox{in vacuum}, r>0)\,\,\,\,\,\,\,\label{genp}
\eea
Which requires either $p=-1$, $p=0$, or $p=2$, which are the solutions given above in Eq.~(\ref{gen}).

Let us stress again that even obtaining the cosmological constant-like, $\kappa_2\,r^2$, correction relies upon the {\em assumptions} of static corrections, spherically symmetric boundary conditions, and ignoring fluctuations around the mean. When including fluctuations and/or relaxing spherical symmetry, the more general form of the potential will be determined in Section \ref{Fluctuations}, finding corrections that appear to be incompatible with observations.

\section{(Non)-MONDian Dynamics}

Let us proceed further. Even though the linear term $\kappa_1\,r$ is forbidden when the equation is solved self-consistently, let us discuss its consequences anyhow, as this was the second focus of Ref.~\cite{Oppenheim:2024rcp}. 

Following Ref.~\cite{Oppenheim:2024rcp}, let us compute the acceleration on scales small enough that the $\kappa_2\,r^2$ is not important; galactic scales. Then we have
\beq
\ddot{\bf x}=-\nabla\Phimpp=-{GM\over r^2}\hat{r}-\kappa_1\hat{r}
\eeq
Neither of these contributions to the acceleration ($1/r^2$ and a constant) look relevant to MONDian dynamics. The basic idea of MOND in Refs.~\cite{Milgrom:1983ca,Milgrom:1983pn,Milgrom:1983zz,Bekenstein:1984tv} is that there is a new contribution to the acceleration which is $\propto \sqrt{M}/r$. It is the $\sqrt{M}/r$ law that is able to reproduce asymptotically flat rotation curves and the Tully-Fisher relation $M\propto v^4$. There is no evidence that an asymptotically constant acceleration would be relevant, as it would produce asymptotic velocity curves {\em growing} with radius as $\sqrt{r}$, rather than flat (this readily follows from considering circular behavior with centripetal acceleration $a=v^2/r$).

To overcome this, in Ref.~\cite{Oppenheim:2024rcp} the manipulation was then to square the above expression
\bea
(\ddot{\bf x})^2&=&\left({GM\over r^2}+\kappa_1\right)^2\\
&=&{G^2M^2\over r^4}+\kappa_1^2+{2GM\kappa_1\over r^2}
\eea
Then by considering the large $r$ region, the first term can be ignored, giving
\beq
(\ddot{\bf x})^2\approx\kappa_1^2+{2GM\kappa_1\over r^2}
\eeq
Then it was indicated that, apart from the constant term, the remaining $2GM\kappa_1/r^2$ term can obtain MOND. 
%the first term here was simply ignored apparently because it is merely a constant, leaving only the remaining $2GM\kappa_1/r^2$ term. 
By taking a square root to recover the acceleration, this appears to give the desired $\propto\sqrt{M}/r$ force law of MONDian dynamics. And the constant pre-factor $\kappa_1$ is to play the role of $a_0/2$ of MOND, where $a_0\sim 10^{-10}\,\mbox{m}/\mbox{s}^2$ is the critical acceleration in which Newton's law transitions from $1/r^2$ to $1/r$. 

However, this procedure is incorrect and is the second problem in the analysis. One cannot take a sum of two terms, $1/r^2$ and constant, square the sum, and note that there is a cross term whose square root has a geometric mean  of the desired $\sqrt{M}/r$ form. 

Instead the acceleration is Newton's $G_NM/r^2$ plus a constant. And there is no evidence that a constant correction helps to reproduce the observed galaxy rotation curves. 
Test particles do not respond to other objects in a way independent of their distance or mass.

Moreover, as stated in the previous section, the constant (arising from the gradient of the linear $\kappa_1\,r$ term) is actually absent when the MNE is solved properly. 
Thus, one in fact only has Newtonian gravity, and one can include a cosmological constant (from $\kappa_2\, r^2$) if desired by imposing static and spherically symmetric boundary conditions.

\section{Fluctuations}\label{Fluctuations}

In an updated version of Ref.~\cite{Oppenheim:2024rcp}, the status of the linear term has been demoted from a vacuum solution (as we showed it is not) to just be a representative possible  {\em fluctuation} from the path integral.
However, while it is true that all paths can contribute to the path integral, a term of the form $\propto r$ is of no more significance than any other power, as it does {\em not} solve the equation of motion. As  can be seen in Eq.~(\ref{genp}), it is as arbitrary as all sorts of other power laws, such as $r^3$ or $r^4$ or $1/r^2$ or $1/r^3$, etc, which do not solve the equation either. 

In fact the situation is even much worse: there is no reason for the fluctuations to be spherically symmetric or even approximately so.
In order to actually study the properties of the fluctuations, we need to return to the probability density function. A careful analysis of this will show a third problem in the analysis.

A general potential configuration can be written as
\beq
\Phi=\Phi_N+\pt
\eeq
where $\Phi_N$ is the usual Newtonian potential obeying the standard Poisson equation $\nabla^2\Phi_N=4\pi G_N\rho$ and $\pt$ is a perturbation. By inserting this into the action of Eq.~(\ref{Action}) we have
\beq
\mathcal{I}= -\alpha\! \int \!dt\, d^3x\left(\nabla^2\pt(x)\right)^2
\eeq
Exponentiating this $\sim e^{\mathcal{I}}$ and integrating gives a probability update rule (see ahead to Eq.~(\ref{ProbUpdate}) for the precise statement of this).

We see that the matter density $\rho$ has dropped out of this. As an application; if there is a black hole, there is no reason for the fluctuation $\pt$ to be spherically symmetric with a singular function $\pt\propto r$ (non-differentiable around $r=0$) as the location of the black hole is not present in this expression. Of course, if there is a non-zero $\pt$ present in the early universe, matter may be attracted to local minima in it, but it will not be exactly at the minimum, nor will it be singular like $r$.

\subsection{Boundary Conditions}

Let us make a note on boundary conditions here. We could go a step further and decompose 
\beq
\pt=\Phi_h+\ptt
\eeq
where $\Phi_h$ obeys $\nabla^4\Phi_h=0$ as introduced earlier. One can use $\Phi_h$ to enforce boundary conditions, while leaving $\ptt$ to obey trivial boundary conditions; $\ptt|_{bdy}=0$ and $\nabla\ptt|_{bdy}=0$. (Under the assumptions of static, spherically symmetric boundary conditions, we can have $\Phi_h=\kappa_2\,r^2$ modulo corrections from $\rho$, as discussed above.)
If we do this, then one can readily use integration by parts to write the action as
\beq
\mathcal{I}= -\alpha\! \int \!dt\, d^3x\left(\left(\nabla^2\Phi_h(x)\right)^2+\left(\nabla^2\ptt(x)\right)^2\right)
\eeq
The fact that there is no linear term in $\ptt$ is precisely what the Euler-Lagrange variation ensures. 
When exponentiated, the first term is just a constant prefactor that implements boundary conditions, while the second term gives a probability distribution rule for fluctuations $\ptt$.
This shows that {\em the probability distribution for fluctuations $\ptt$ are in fact uncorrelated with $\Phi_h$} (such as $\kappa_2\,r^2$.)

In Ref.~\cite{Oppenheim:2024rcp} even the $\kappa_2\,r^2$ gets treated probabilistically and therefore it is not really a fixed boundary condition. In this case, we should actually just return to $\pt$ as the generic form of any fluctuation about the Newtonian potential $\Phi_N$ and study the full distribution.

\subsection{Full Distribution}\label{FullDistn}

Let us now examine in some detail the actual distribution. Often probability distributions can have interesting temporal dependence through the $dt$ integral, giving a rule for how to update the distribution from an initial time $t_i$ to a final time $t_f$, as
\beq
\varrho(\phi;t_f)=\mathcal{N}\!\int\! \mathcal{D}\phi \, \exp\left[-\alpha\int_{t_i}^{t_f} dt\,d^3x(\nabla^2\phi(x))^2\right]\varrho(\phi;t_i)
\label{ProbUpdate}\eeq
where $\mathcal{N}$ is a normalization constant.
However, in Ref.~\cite{Oppenheim:2024rcp} it is suggested that it should be static in this Newtonian regime. As mentioned earlier, one may anticipate important temporal variation on the Hubble time as the universe expands or on other dynamical time scales. So a static assumption is not clearly justified. 

In fact as written the form presented is not well defined, as there are no time derivatives in this action. This means that in the path integral each moment in time is {\em decoupled} from the others. By breaking up the integral over time into a Riemann sum $\int_{t_i}^{t_f}dt\to\epsilon\sum_i^f$, where $\epsilon=dt$ is the time step, we see that the probability distribution {\em factorizes} and so all earlier times become unimportant. In the continuum $\epsilon=dt\to0$ limit, we then have an un-normalized distribution, unless one sends the factor $\alpha\to\infty$ to compensate. If one does this, or if one simply introduces a hard cut off in time ($\epsilon$ remains finite), then one can normalize the distribution, but one should expect the distribution to jump around in time in an uncorrelated fashion.
It could be that when one includes the contribution to the path integral from the (quantum) matter degrees of freedom, the situation is altered; we do not develop this issue further here. But we do consider relativistic corrections in Section \ref{Temporal}, which can provide time derivatives. 

For now, we shall proceed as is done in Ref.~\cite{Oppenheim:2024rcp} by
ignoring the $dt$ integral and the temporal dependence. If there are no prior fixed boundary conditions, then the probability distribution for a fluctuation $\pt$ at any moment in time is the Gaussian distribution
\beq
\varrho(\pt)=\mathcal{\tilde{N}}\exp\left[-\alpha_T\int d^3x(\nabla^2\pt)^2\right]
\label{distn}\eeq
where $\alpha_T\sim\alpha\,\epsilon$ has units of length (we set $c=1$ here; if we reinstate factors of $c$, it has units of time$^4$/length$^3$).  

We note that technically one must implement some boundary conditions on $\phi$, or otherwise this distribution is not normalizable, since if we shift $\pt\to\pt+\psi_h$, where $\nabla^2\psi_h=0$ obeys the Laplace equation, there is no change in probability. 
If boundary conditions are enforced on the scale of our horizon, it can have an impact on comparable scales, but should not be relevant on the scale of individual galaxies as they are orders of magnitude smaller than the Hubble scale.
However if we imagine implementing the boundary conditions on a scale much larger than our horizon, there should be no change in the bulk distribution, and so these details will be unimportant. We shall assume this simpler setup in the following. 

\subsection{Power Spectrum}

It is convenient to switch to Fourier space, giving
\beq
p(\pt)=\mathcal{\tilde{N}}\exp\left[-\alpha_T\int {d^3k\over(2\pi)^3}\,k^4|\pt_k|^2\right]
\eeq
Such a Gaussian distribution is characterized by its 2-point correlation function
\beq
\langle\pt_k\,\pt^{*}_{k'}\rangle=(2\pi)^3\delta^3({\bf k}-{\bf k}')\,P_\pt(k)
\eeq
where the {\em power spectrum} $P_\pt(k)$ is read off to be
\beq
P_\pt(k)={1\over 2\,\alpha_T\,k^4}
\label{spectrum}\eeq
We emphasize this spectrum necessarily follows from the set-up laid out in Ref.~\cite{Oppenheim:2024rcp} (although the static/non-relativistic treatment is questionable and will be addressed in the next Section.) 
We also note that since this is all derived in kind of Newtonian approximation, we only expect it to apply on sub-horizon scales, i.e., $k\gtrsim H_0$ the Hubble constant.

In position space, the corresponding 2-point correlation function is
\bea
\langle\pt({\bf x})\,\pt({\bf y})\rangle &=&\int{d^3k\over(2\pi)^3}e^{i{\bf k}\cdot({\bf x}-{\bf y})}P_\pt(k)\\
&=&\int d\ln k\,{\sin(k L)\over k L}{k^3\,P_\pt(k)\over 2\pi^2}
\eea
where $L=|{\bf x}-{\bf y}|$ is the distance between 2 points of interest. Here the lower end of the $k$ integral should be cut off at $k\sim H_0$, while the upper end of the integral doesn't obviously need to be cut off since it is sufficiently UV soft (we shall revisit this when studying the acceleration below). 

To get some intuition for this, we see that the characteristic fluctuation in $\phi$ on a scale $k\sim1/L$ is (the standard deviation per log interval)
\beq
\sigma_k=\sqrt{k^3\,P_\pt(k)\over 2\pi^2} = {1\over 2\pi\sqrt{\alpha_T\,k}}
\eeq
In Ref.~\cite{Oppenheim:2024rcp} the value of $\alpha_T$ is taken to be 
\beq
\alpha_T\sim 0.01/\sqrt{\Lambda}\sim0.01/H_0
\eeq
This was selected to ensure that the variance of $\kappa_2$ is of the right order of magnitude. (This follows from setting $\phi=\kappa_2\,r^2$, inserting into the above distribution (\ref{distn}) to obtain $\varrho(\pt)\propto \exp(-36\,\alpha_T\,V\,\kappa_2^2)$, with volume $V=(4\pi/3)H_0^{-3}$, giving $\langle\kappa_2^2\rangle=H_0^3/(96\pi\alpha_T)$. By demanding the standard deviation is of the order of the observed $\kappa_2=-\Lambda/6$ and using the fact that 
the observed cosmological constant is a significant fraction of the energy of the present universe, $\Lambda\sim 3H_0^2$, we obtain the above $\alpha_T$).

\subsubsection{Dark Energy Behavior}

In this full analysis, we see that this corresponds to having fluctuations be $\mathcal{O}(1)$ on the scale of the horizon $k\sim H_0$. This allows one to try to claim that one has a kind of dark energy. However, we see here that there is no reason for such a fluctuation to be spherically symmetric, or precisely of the quadratic form $\kappa_2\,r^2$, unless one imposes this constraint by hand. Hence one is not in fact actually recovering a kind of cosmological constant as a likely fluctuation. In stark contrast, $\mathcal{O}(1)$ fluctuations on the scale of the horizon are more statistically likely to lead to black hole formation.

\subsubsection{Large Scale Structure}

\begin{figure}[t!]
\centering
\includegraphics[width=\columnwidth]{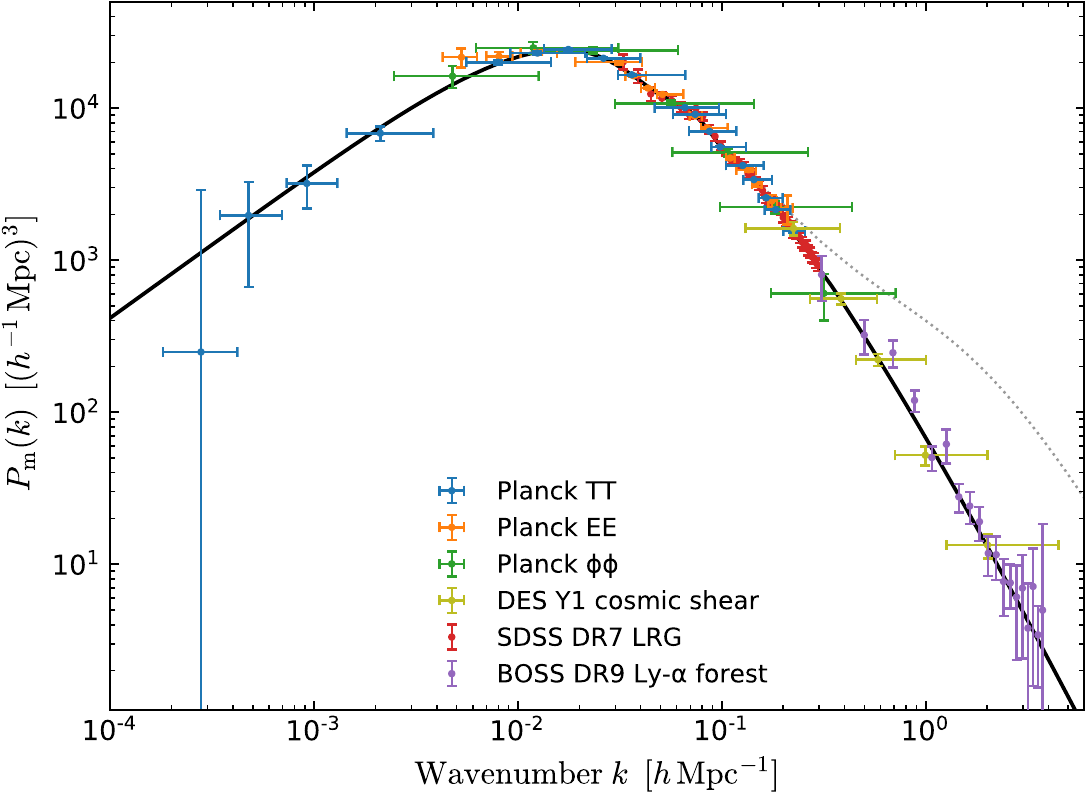}
\caption{Observed power spectrum of fluctuations from a range of observations extracted to today.
This Figure is taken from Ref.~\cite{Planck:2018nkj}.  
This can be understood as the gravitational potential $\Phi$ fluctuations, but with a re-scaled factor, according to Eq.~(\ref{Prescale}). By comparison, the PQCG makes a prediction of a flat spectrum, whose amplitude is controlled by the parameter $\alpha_T$, as given in Eq.~(\ref{Pflat}). Both the amplitude (for anything near the suggested $\alpha_T\sim 0.01/H_0$) and the shape are clearly incompatible with observations.}
\label{FigPowerSpectrum} 
\end{figure}

Moreover, on sub-horizon scales, we can test if this spectrum of fluctuations is compatible with observations. The concordance model (CM) in cosmology with baryons, dark matter, and dark energy has a spectrum of fluctuations for $\Phi=\Phi_N$ (deterministic, but arising from 
$\nabla^2\Phi_N/a_s^2=4\pi G\bar{\rho}_m\delta_m=3 H^2\Omega_m\delta_m/2$ due to
inhomogeneous matter $\delta\rho_m=\bar{\rho}_m\,\delta_m$ drawn from some distribution). This is known to be nicely compatible with observations. In linear theory, it takes on the form (for a review, see Ref.~\cite{Hertzberg:2012qn})
\beq
P_{\CM}(k,a_s)={9\pi^2\over 2}\delta_H^2\,\Omega_{m,0}^2{k^{n_s-4}\over H_0^{n_s-1}}
T^2(k)\left(D(a_s)\over a_s\,D(1)\right)^2
\eeq
where $a_s$ is the scale factor with $a_s=1$ today, $D(a_s)$ is the so-called growth factor, and $\Omega_{m,0}\approx0.25$ is today's fraction of matter in the CM. 
The overall amplitude of fluctuations $\delta_H$ and the spectra tilt $n_s$ are measured to be
\beq
\delta_H\approx 5\times 10^{-5},\,\,\,\,\,n_s\approx 1
\eeq
(with $n_s=0.96 - 0.97$ the more precise value).
The so-called transfer function takes on the approximate asymptotic forms
\beq
T(k)=\Bigg{\{}
\begin{array}{l}
1,\,\,\,\,k\lesssim k_{eq}\\
{12k_{eq}^2\over k^2}\,\ln\left(k\over 8 k_{eq}\right),\,\,\,\,k\gg k_{eq}
\end{array}
\eeq
(up to wiggles from baryon-acoustic-oscillations)
where the break is provided by the scale of matter-radiation equality
\beq
k_{eq}\approx 0.073 \,\mbox{Mpc}^{-1}\Omega_{m,0}h^2
\eeq
Using the fact that $n_s$ is close to 1, we have the spectrum today $a_s=1$ of
\beq
P_{\CM}(k)\sim 10^{-8}\,k^{-3}\,T^2(k)
\label{spectrumCM}\eeq
Importantly, this spectrum is consistent with a range of cosmological surveys (for example, see Section 4 of Ref.~\cite{Planck:2018nkj} for a review). By using the fact that baryons respond to the gravitational potential through ${\bf a}=-\nabla\Phi$ in the CM, one obtains the observed spectrum given in Figure \ref{FigPowerSpectrum}. The spectrum plotted is not quite $P_{\CM}$, but a re-scaled version given by
\beq
P_{m,\CM}(k)={4\over 9\,\Omega_{m,0}^2\,H_0^4}\,k^4\,P_{\CM}(k)
\label{Prescale}\eeq
On the other hand, by performing this re-scaling of PQCG in eq.~(\ref{spectrum}), one has
\beq
P_{m,\phi}(k)={2\over 9\,\Omega_{m,0}^2\,H_0^4\,\alpha_T}\approx 10^{13}\!\left(0.01/H_0\over\alpha_T\right)\!(h^{-1}\mbox{Mpc})^3
\label{Pflat}\eeq
i.e., a flat spectrum. (In the CM, $P_{m,\CM}$ is physically interpreted as the ``matter power spectrum", while in PQCG it does not directly have this interpretation as the fluctuations $\phi$ have no source. But what we observe are the effects of the gravitational potential, and so this factor of ($4 k^4/(9\Omega_{m,0}^2H_0^4)$ can be viewed as just a convenient re-scaling. However, in this case the mean about which one is expanding only arises from the baryons, as the fluctuations $\phi$ have vanishing mean. Having the mean all come from baryons is already very strong indication that such a framework cannot match cosmological data and provide a flat universe, etc, but the details of the fluctuations are highly problematic, as we focus on here. 

We see that the prediction from PQCG of a flat power spectrum in the $P_m(k)$ variable is very different from the observed power in Figure \ref{FigPowerSpectrum}. By comparing Eqs.~(\ref{spectrum}) and (\ref{spectrumCM}), if $\alpha_T\sim 0.01/H_0$ as chosen in Ref.~\cite{Oppenheim:2024rcp}, then the power in PQCG is $\approx10$ orders of magnitude too large on the scale of the horizon. And furthermore, it remains far too large even for modes that are several orders of magnitude within the horizon. This spectrum is clearly ruled out by observations.

To get some intuition for its impact, consider a test particle (baryon) subject to this new acceleration: $a=|\nabla\phi|$. The typical value for this on a scale $L$ is 
\beq
a\sim\sigma_k/L\sim\sqrt{k/\alpha_T}/(2\pi)
\label{accest}\eeq
We should compare this to the characteristic acceleration normally experienced by a particle in a standard FLRW universe from matter
\beq
a_{\mbox{\tiny{FLRW}}}={GM_{enc}\over L^2}= {1\over2}\Omega_mH^2L\sim H^2/k
\eeq
(in a matter era). 
The typical scales on which a region collapses away from the cosmic expansion is when $a\gtrsim a_{\mbox{\tiny{FLRW}}}$. By comparing these last 2 expressions at the present epoch $H=H_0$, we see that if $\alpha_T\sim0.01/H_0$, then collapse can occur for all sub-horizon modes $k\gtrsim H_0$. 
Thus leading to a radically altered universe, looking nothing at all like ours.

\subsubsection{Galactic Behavior}

Alternatively, one could try to avoid these huge fluctuations on large scales by increasing $\alpha_T$ by at least 10 orders of magnitude to be
\beq
\alpha_T\gtrsim 10^8/H_0
\eeq
so that these fluctuations are no larger than the observed fluctuations on the scale of the horizon today. 
However this leads to a spectrum that is then much lower than the observed spectrum on sub-horizon scales due to the faster fall off ($1/k^4$ versus $1/k^3$). For $k\gg k_{eq}$ it does not have the observed  break in the spectrum (and could then potentially be too large depending on $\alpha_T$). By comparing the flat prediction of PQCG in Eq.~(\ref{Pflat}) to the observed spectrum of Figure \ref{FigPowerSpectrum}, we see that it is clearly different. 

Moreover, it is also much lower on galactic scales, since the actual spectrum is enhanced relative to the linear theory summarized above due to nonlinear dynamics. In typical halos like the milky way $\Phi\sim 10^{-6}$, while this theory would then be bounded by $\sigma_k\lesssim10^{-5}\sqrt{H_0/k}$; so for $k\gg 10^2H_0$ (which is true for known galaxies whose flattened rotation curves start around $\sim 10$\,kpc) then $\sigma_k\ll \Phi$. For known galaxies whose flattened rotation curves start around $r\sim 10$\,kpc, one might estimate $k\sim 2\pi/(4\times10\,\mbox{kpc})$ (so that $r\sim10$\,kpc corresponds to a quarter wavelength), then $k/H_0\sim 10^6$, giving $\sigma_k/\Phi\lesssim 10^{-2}$. 
Thus making these fluctuations too small to directly explain galactic rotation curves. 

Also, for these large values of $\alpha_T$, it means that this mechanism would be completely irrelevant for mimicking dark energy as a stochastic fluctuation. Dark energy would then have to merely arise from an imposed boundary condition provided by $\Phi_h$ and the statistical fluctuations in this setup become cosmologically irrelevant. 

\subsubsection{Small Scale Behavior}

On much smaller scales, there is another problem. The 2-point correction function for a test particle's (baryon) acceleration ${\bf a}=-\nabla\phi$ is
\bea
\langle{\bf a}({\bf x})\cdot{\bf a}({\bf y})\rangle 
&=&\int d\ln k\,{\sin(k L)\over k L}{k^3\,P_a(k)\over 2\pi^2}\\
&=&{1\over 8\pi\alpha_T L}\label{acc2}
\eea
where we used $P_a(k)=k^2\,P_\pt(k)=1/(2\alpha_T k^2)$. This is just the more precise version of the earlier estimate in Eq.~(\ref{accest}). (Here the infrared part of the integral is soft, so we extended the integral to $k\to 0$.) We note that if we were to compute the variance in a particle's acceleration by taking ${\bf y}\to {\bf x}$ (i.e., $L\to 0$) this diverges. Hence this theory is again not well defined. However, one can imagine a cut off $k_{\mbox{\tiny{UV}}}$ on the UV k-modes to regulate this (perhaps near the Planck scale or so). Relatedly, one can consider the physically important quantity of relative accelerations $\langle({\bf a}({\bf x})-{\bf a}({\bf y}))^2\rangle$ and regulate accordingly. 

For a collection of nearby particles, separated by a scale $L$ above the cut off, this formula tells us that their stochastic relative acceleration can be quite large. For earth based Cavendish-type tests of gravitation, the acceleration is suppressed by a factor of $M$ the mass of the source. However, in this theory there is no suppression by the mass of the source. So even though this is a contribution to the acceleration that on small distances, only rises as $a\propto 1/\sqrt{L}$, rather than Newton's inverse square law, the fact that there is no $M$ suppression means it can be relatively large on small scales. 

For example, in Ref.~\cite{Hoyle:2004cw} objects of mass $M\sim 30$\,gm at a distance of $r\sim 30$\,mm have been measured and found to agree with Newton's law to good precision (there are multiple updates to smaller distances, but we take this as an informative starting point). This is an acceleration of 
\beq
a_N={GM\over r^2}\approx2\times10^{-9}\,\mbox{m/s}^2
\eeq
On the other hand, if we consider the stochastic contribution above on the same scale $L=30$\,mm, we have
\beq
\sqrt{\langle{\bf a}\cdot{\bf a}\rangle}_{L=30\,mm}\approx{9\times10^3\over\sqrt{\alpha_T\,H_0}}\,\mbox{m/s}^2
\eeq
Hence this is orders of magnitude too large.
In order for this new, as yet unobserved, stochastic contribution to be smaller than the observed acceleration, we have a much tighter bound on $\alpha_T$ of
\beq
\alpha_T\gtrsim 2\times 10^{25}/H_0
\eeq
For these extremely large values of $\alpha_T$ (compare to the $\alpha_T\sim 0.01/H_0$ chosen in Ref.~\cite{Oppenheim:2024rcp}) the effects on cosmological scales are completely irrelevant as the power spectrum is reduced by some 27 orders of magnitude. 

Perhaps by including temporal stochastic behavior (see next section) the net effect on acceleration will be reduced, thus allowing for smaller values of $\alpha_T$; but this is highly unlikely to change this conclusion so drastically that anything close to $\alpha_T\sim 0.01/H_0$ becomes allowed.

An alternative to avoid this conclusion would be to lower the UV cut off so much that any sub-galactic fluctuations are suppressed (i.e., take $k_{\mbox{\tiny{UV}}}\sim k_{\mbox{\tiny{galaxy}}}$). But then such a theory of gravity is not useful over a wide range of scales (table-top, solar system, etc) for which it is already well studied, and the UV problems of quantum gravity are not addressed at all.

%\bigskip
\section{Temporal Stochasticity}\label{Temporal}

The temporal dependence of the fluctuations is something that should also be carefully addressed. As mentioned earlier, the theory may conceivably lead to changes on the Hubble time or other dynamical times. In fact, as discussed earlier, the path integral suggests that the distribution should in fact jump significantly from one time step to the next. This could lead to altered constraints in the problems identified above. On the one hand, some constraints could be weakened due to temporal variation getting partially washed out through temporal averaging. On the other hand, some constraints could be be strengthened due to a new kind of, as yet unseen, temporal jitter in the behavior of gravity. All this deserves careful consideration.

\subsection{Relativistic Corrections}

As a step in this direction, let us reinstate relativistic corrections. The full action proposed is
\bea
\mathcal{I}=-\hat{\alpha}\int dt\,d^3x\sqrt{-g}\Big{[}(G_{\mu\nu}-8\pi G_N T_{\mu\nu})^2\nonumber\\
\,\,\,-\beta(G-8\pi G_NT)^2\Big{]}
\eea
where $G_{\mu\nu}$ is the Einstein tensor, $T_{\mu\nu}$ is the energy-momentum tensor, and $\hat{\alpha},\,\beta$ are constants.
To study the fluctuations, we can expand around the solution of Einstein's equations as
\beq
g_{\mu\nu}=g_{\mu\nu,E}+h_{\mu\nu}
\eeq
where $g_{\mu\nu,E}$ obeys the Einstein field equations $G_{\mu\nu,E}=8\pi G_N T_{\mu\nu}$ and $h_{\mu\nu}$ is a fluctuation. Let us consider scalar fluctuations in the metric as
\beq
h_{\mu\nu}=2\,\phi\,\delta_{\mu\nu}
\eeq
Then working to quadratic order, we obtain
\bea
\mathcal{I}=-\alpha\int dt\,d^3x\Big{[}(\nabla^2\phi)^2-2b\,\nabla^2\phi\,\ddot\phi/3
+b\,(\ddot\phi)^2\Big{]}
\eea
where $\alpha=4(1-\beta)\hat{\alpha},\,b=3(1-3\beta)/(1-\beta)$. Here one demands $\beta<1/3$ (or the stronger constraint $\beta\leq0$) for positivity of $b$, along with $\alpha>0$. Note that one picks up time derivatives here; such terms are in fact suppressed by factors of $1/c$ if we reinstate units by replacing $\ddot\phi\to\ddot\phi/c^2$. This means that when inserted into the path integral, it is better behaved in the sense that the distribution will not jump around at every instant.

By passing to Fourier space, we can compute the 2-point correlation function in time as
\beq
\langle\phi_k(t)\,\phi^*_{k'}(t')\rangle=(2\pi)^3\delta^3({\bf k}-{\bf k}')Q_\phi(k;t,t')
\eeq
where the ``power spectrum" $Q$, including temporal correlations, is
\beq
Q_\phi(k;t,t')=\int{d\omega\over(2\pi)}{e^{-i\omega (t-t')}\over \alpha(k^4-2b\,k^2\,\omega^2/3+b\,\omega^4)}
\eeq
We note that for $b>0$ there are no poles along the real $\omega$ line; this corresponds to correlations being {\em exponentially suppressed in time}. 

This integral can be carried out using the residue theorem. The full details are not so important, but the qualitative structure is
\beq
Q_\phi(k;t,t') = {e^{-a\,c\,k\,|t-t'|}\over\alpha\,k^3}\,f(a\,c\,k\,|t-t'|)
\eeq
where $a>0$ is an $\mathcal{O}(1)$ number (assuming $\mathcal{O}(1)$ values of $\beta$) and $f$ is a periodic function with period 1 and an $\mathcal{O}(1)$ amplitude. We have instated a factor of $c$ into the exponent to highlight the role that $c$ is playing here. 
the exponential factor shows that (up to an $\mathcal{O}(1)$ factor) there is only temporal correlation for a period of time 
\beq
T_k={1\over a\,c\,k}
\eeq 
for the mode of interest. If we were to take the $c\to\infty$ limit, i.e., the Newtonian limit, this becomes
\beq
Q_\phi(k;t,t') \sim {1\over\alpha\,k^4}\delta(t-t')
\eeq
i.e., the fluctuations become uncorrelated in time in this Newtonian limit, as we already discussed below Eq.~(\ref{ProbUpdate}). At the equal time $t=t'$ the power spectrum 
\beq
P_\phi(k)=Q_\phi(k;t,t)
\eeq
would be formally infinite as this is where the delta-function hits. This again reinforces the points made earlier, as we defined a new parameter $\alpha_T\sim \alpha\,\epsilon$ where $\epsilon$ was some temporal cut off, to give the power $P\sim 1/(\alpha_T\,k^4)$. These steps were needed to obtain the kind of static analysis of Ref.~\cite{Oppenheim:2024rcp} who ignored the temporal integral. 

By keeping $\alpha$ a finite parameter of the theory and properly tracking the time dependence, we see that relativistic effects regulate the temporal correlation function then away from a delta-function to the exponential factor. We can think of this as the replacement
\beq
\delta(t-t')\to {a\,c\,k\over2}e^{-a\,c\,k|t-t'|}
\eeq
as these 2 functions have the same form in the $c\to\infty$ limit. 

At equal times in the relativistic theory, we have
\beq
P_\phi(k)={f(0)\over\alpha\,k^3}
\eeq
with $f(0)$ an $\mathcal{O}(1)$ number. So one can draw the fluctuations from this scale invariant spectrum at a given moment in time, but bearing in mind that they will deviate away from this by an $\mathcal{O}(1)$ amount on the time scale $T_k \sim 1/(c\,k)$, the light crossing time over a wavelength. By considering the standard deviation per log interval it is now
\beq
\sigma_k=\sqrt{k^3\,P_\pt(k)\over 2\pi^2} \sim {0.1\over\sqrt{\alpha}}
\eeq

So to obtain a kind of dark energy like contribution, one would take 
\beq
\alpha\sim 0.01
\eeq
so that fluctuation are $\mathcal{O}(1)$ on the scale of the horizon
(this is the analogue of taking $\alpha_T\sim 0.01/H_0$ in Ref.~\cite{Oppenheim:2024rcp} that we described earlier, when the temporal integral was ignored).  For these horizon scale modes, the correlation time-scale $T_k\sim 1/(c\,k)$ is of order the Hubble time. So the value of this putative dark energy would fluctuate on the Hubble time, potentially changing from positive to negative, etc. Such behavior is not supported by existing data.

More problematically,  this also means the spectrum is $\approx 10$ orders of magnitude too large for sub-horizon scales; compare to the observed spectrum in Eq.~(\ref{spectrumCM}). Again this is clearly ruled out. 

If one raises $\alpha$ considerably to $\alpha\sim 10^{8}$, then while a scale invariant spectrum may at first sight seem promising, it
does not have the observed break in the spectrum at matter-radiation equality; compare to  Figure \ref{FigPowerSpectrum} for the related variable $P_m=
4 k^4/(9\Omega_{m,0}^2H_0^4)\,P_\phi(k)$. 

On galactic scales, a new problem is that these temporal correlations mean that there is a radical change in the gravitational field on the order the light crossing time $T_k\sim 1/(c\,k)$. So therefore even if stars were to be orbiting in this stochastic gravitational field that Ref.~\cite{Oppenheim:2024rcp} proposes replaces dark matter and provides the galactic rotation curves, the stars orbits would drastically change on a light crossing time. So for example, for flattened rotation curves starting at $r=10\,$kpc, after $\approx 30,000$ years the stars would likely start orbiting in a completing different direction as the gravitational field has completely changed. This temporal change in $\Phi$ would likely disrupt halos as there would just be an incoherent mess in the gravitational potential on times longer than any light crossing times.

Finally, let us consider small scale experiments. If we consider the 2-point correlation function for acceleration of test particles, and for simplicity treat the $f$ as slow, we have
\bea
\langle{\bf a}({\bf x},t)\cdot{\bf a}({\bf y},t')\rangle 
&=&\int d\ln k\,{\sin(k L)\over k L}{k^3\,Q_a(k,t,t')\over 2\pi^2}\,\,\,\,\,\\
&\sim&{f(0)\over 2\pi^2\alpha (L^2+a^2\,c^2\,(t-t')^2)}
\eea
(in fact there is an $\mathcal{O}(1)$ correction from the details of $f$, but that it is not essential here). The physically important quantity of relative accelerations $\langle({\bf a}({\bf x},t)-{\bf a}({\bf y},t'))^2\rangle$ is a simple extension of the discussion here. 
Note that if we now consider the equal-time 2-point correlation function it rises with small $L$ as $1/L^2$, rather than just $1/L$ as seen earlier in Eq.~(\ref{acc2}). By again considering a table-top Cavendish experiment at $L=30$\,mm, we have
\beq
\sqrt{\langle{\bf a}\cdot{\bf a}\rangle}_{L=30\,mm}\approx{7\times10^{17}\sqrt{f(0)}\over\sqrt{\alpha(1+\tau^2})}\,\mbox{m/s}^2
\eeq
where $\tau=a^2c^2(t-t')^2/L^2$. At a given moment in time $\tau=0$, this suggests that to avoid being larger than the observed value of $\sim 2\times 10^{-9}$m/s$^2$, one requires the extraordinarily large value $\alpha\gtrsim 10^{53}$. However, since the experiment takes place over a period of time much longer than the light crossing time, one should consider the factor that there is some temporal suppression through the above $\tau$ factor. However the temporal suppression here is only power law, with $\sqrt{\langle{\bf a}\cdot{\bf a}\rangle}\propto 1/\tau$ at times longer than the light crossing time. So while the residual bound on $\alpha$ may be several orders of magnitude lower than $10^{53}$ it is highly unlikely to be anywhere near the kinds of values, like $\alpha\sim 0.01$ for non-trivial cosmological consequences to be allowed.

\subsection{Other Works}
In addition, there have been other works, such as Ref.~\cite{deCesare:2016dnp}, explicitly suggesting that aspects of gravitational theory, such as Newton's constant $G_N$,  vary in time stochastically in order to give rise to dark energy. Such a proposal could be ruled out by lunar ranging measurements, which constrain $G_N$ to change by at most a small amount on the Hubble time \cite{Muller:2007zzb}.

\section{Negative Densities}

Let us also note that these fluctuations $\phi$ are associated with a kind of so--called ``phantom dark matter $\rho_{ph}$" defined through $\nabla^2\phi=4\pi G_N\rho_{ph}$. But since $\phi$ is drawn from a Gaussian with zero mean, then so too is $\rho_{ph}$ (albeit with an altered, white noise, spectrum). So this type of ``phantom dark matter" density would randomly fluctuate from place to place with both positive and negative values (and zero mean). Again there is no evidence whatsoever that dark matter can be modeled this way. In particular, the existence of regions of space with negative densities has no known supporting evidence in its favor at present. 

In fact we can even go further: in the halos and between galaxies, where the baryons are negligible, this ``phantom density" would dominate, causing the total density to fluctuate to negative values in places. As is well known, negative densities lead to violations of the null energy condition (NEC), which in turn causes signals to become superluminal (as opposed to the Shapiro time delay, one has Shapiro time advance with NEC violation). This leads to possible acausality. This indicates a fundamental problem inherent in these constructions.

%\bigskip

\section*{Acknowledgments}
%{\em Acknowledgments}.---
We thank Mordehai Milgrom for helpful discussion. We thank Jonathan Oppenheim and Andrea Russo for discussion. 
M.~P.~H.~ is supported in part by National Science Foundation grants PHY-2310572 and PHY-2013953. 
A.~L.~ was supported in part by the Black Hole Initiative at Harvard University which is funded by grants from the John Templeton Foundation and the Gordon and Betty Moore Foundation.

%\appendix

\end{document}